\documentclass{article}

\usepackage{arxiv_template}

\usepackage[utf8]{inputenc} 
\usepackage[T1]{fontenc}    
\usepackage[hyperfootnotes=False]{hyperref}       
\usepackage{url}            
\usepackage{booktabs}       
\usepackage{amsfonts}       
\usepackage{nicefrac}       
\usepackage{microtype}      
\usepackage{fancyhdr}       
\usepackage{graphicx}       
\usepackage{amsmath}
\usepackage{algorithm, algorithmicx, algpseudocode}
\usepackage{threeparttable}
\usepackage{multicol}

\usepackage{subcaption}
\usepackage[labelsep=quad,labelfont=bf]{caption}

\usepackage{graphicx}

\newcommand{\grey}[1]{\textcolor[rgb]{0.5,0.5,0.5}{#1}}

\usepackage{color}

\renewcommand{\eqref}[1]{Eq.~(\ref{#1})}
\newcommand{\figref}[1]{Fig.~\ref{#1}}
\newcommand{\tabref}[1]{tab.~\ref{#1}}
\newcommand{\secref}[1]{Sec.~\ref{#1}}

\hypersetup{
    pdfauthor={Siedler TM and Jakob PM and Herold V},
	pdftitle={Enhancing quality and speed in database-free neural network reconstructions of undersampled MRI with SCAMPI},
	pdfborder = {0 0 .2}
}

\fancypagestyle{submissioninfo}{
	\fancyhf{}
	
	\fancyhead[C]{
		{\Large{\grey{This article has been accepted for publication at\\
					 Magn Reson Med - DOI: \href{https://onlinelibrary.wiley.com/doi/10.1002/mrm.30114}{10.1002/mrm.30114}}}		
		}
		}
}

\author{
  Thomas M. Siedler, 
  Peter M. Jakob, 
  Volker Herold \\
  Experimental Physics 5, University of Würzburg, \\
  Am Hubland, 97074 Würzburg \\
  \texttt{volker.herold@physik.uni-wuerzburg.de} \\
}

\begin{document}

\title{Enhancing quality and speed in database-free neural network reconstructions of undersampled MRI with SCAMPI}

\twocolumn[{\csname @twocolumnfalse\endcsname               
    \maketitle
    
    \begin{abstract}
        \textbf{Purpose:}       
		We present SCAMPI (Sparsity Constrained Application of deep Magnetic resonance Priors for Image reconstruction), an untrained deep Neural Network for MRI reconstruction without previous training on datasets. It expands the Deep Image Prior approach with a multi-domain, sparsity-enforcing loss function to achieve higher image quality at a faster convergence speed than previously reported methods.
        \textbf{Methods:}
        Two-dimensional MRI data from the FastMRI dataset with Cartesian undersampling in phase-encoding direction were reconstructed for different acceleration rates for single coil and multicoil data.
        \textbf{Results:} 
        The performance of our architecture was compared to state-of-the-art Compressed Sensing methods and ConvDecoder, another untrained Neural Network for 2D MRI reconstruction. SCAMPI outperforms these by better reducing undersampling artifacts and yielding lower error metrics in multicoil imaging. In comparison to ConvDecoder, the U-Net architecture combined with an elaborated loss-function allows for much faster convergence at higher image quality. SCAMPI can reconstruct multicoil data without explicit knowledge of coil sensitivity profiles.
        Moreover, it is a novel tool for reconstructing undersampled single coil k-space data.
        \textbf{Conclusion:}
        Our approach avoids overfitting to dataset features, that can occur in Neural Networks trained on databases, because the network parameters are tuned only on the reconstruction data. It allows better results and faster reconstruction than the baseline untrained Neural Network approach.
        
        \medskip
        
    
    \end{abstract}
    \vspace{1cm}
    
}]

\let\thefootnote\relax\footnote{\textbf{Abbreviations:} 
BART, Berkeley Advanced Reconstruction Toolbox; CS, Compressed Sensing; DIP, Deep Image Prior; GT, Ground Truth; L1W, Wavelet L1~Norm; MRI, Magnetic Resonance Imaging; MSE, Mean Squared Error; PI, Parallel Imaging; PICS, Parallel Imaging combined with Compressed Sensing; RAKI, Scan‐specific robust artificial‐neural‐networks for k‐space interpolation; SCAMPI, Sparsity Constrained Application of deep Magnetic resonance Priors for Image reconstruction; SNR, Signal-to-noise ratio; SSIM, Structural Similarity Index; TV, Total Variation}

\section{Introduction}\label{sec:introduction}

Increasing acquisition speed is one major goal of ongoing research in Magnetic Resonance (MR) Imaging.
Two common approaches are widely used in the clinical routine to accelerate data acquisition. 
Both are based on extensive undersampling of the MR signal in the Fourier domain.
The first is Parallel Imaging (PI), which exploits the redundancy of data received by several radio-frequency coils
\cite{Griswold.2002,sodickson1997simultaneous,pruessmann1999sense,Deshmane.2012}. 
It can be applied to 2D and 3D imaging and can be combined with techniques like simultaneous multislice excitation \cite{Barth.2016}.
The second one introduces sparsity regularization into the image reconstruction process and is known as
Compressed Sensing (CS)~\cite{lustig2007sparse}. In combination (PICS), both techniques can achieve high acceleration rates \cite{Otazo.2012.Combination,liang2009accelerating}. 
Nonetheless, the reconstruction task in general is an ill-posed optimization problem. Solving it often requires computationally intense iterative algorithms, which limits its application, for example in fields such as real time imaging.
Thus, the need for fast image reconstruction has been one reason for the advent of Neural Networks and deep learning strategies in the application of MRI reconstruction \cite{Lundervold.2019,Liang.2020, Ahishakiye.2021}.

On a high abstraction level, two classes of methods that use Neural Networks can be distinguished. Data-driven machine learning algorithms require large datasets in order to tune network parameters,
so that they `learn' to reconstruct high-quality images from undersampled measurements. Several studies have shown that deep learning approaches outperform the aforementioned state-of-the-art
methods, not only in terms of reconstruction speed, but also in terms of image quality \cite{Schlemper.2018, Knoll.2020.DeepLearning}.
Despite remarkable results, data-based approaches have serious limitations when generalizing from training data: Overfitting to particular features of the training data set can lead
to bias and limit generalization when posed with previously unseen data in new applications. Upon inference on new data, this can lead to hallucination of features that are common in the training data set~\cite{Bhadra.2021.Hallucinations}. Vice versa, it may cause removal of actual features that were never seen before in the training data, which can be fatal for diagnostic imaging \cite{Knoll.2020.DeepLearning}.
Moreover, reconstruction quality of a well-trained Neural Network deteriorates when confronted with new imaging properties,
such as different sampling patterns or signal-to-noise ratio of the underlying data \cite{Knoll.2019.Assessment, Antun.2020.Instabilities}.

Additionally, high quality ground truth data can be scarce or physically impossible to acquire. 
Therefore, various strategies
of unsupervised learning have emerged, that do not require fully sampled reference data, but can be trained on datasets of undersampled measurements \cite{Akcakaya.2022,Liu.2020,Aggarwal.2022,Yurt.2022}.
However, they do not change the general strategy to infer image properties from a training dataset to a new reconstruction task.

The second approach that uses Neural Network algorithms does not require training databases and fits the model parameters only to the individual scan's data \cite{PourYazdanpanah.2019}.
For example, the
RAKI \cite{Akcakaya.2019} architecture extends GRAPPA's \cite{Griswold.2002} linear convolution kernels by multiple layers of scan-specifically trained convolutions, linked by rectified linear units
(ReLU).
The authors point out that non-linear functions with few degrees of freedom are better at approximating higher-dimensional linear functions and show that RAKI has superior
noise-resilience and better reconstruction performance than GRAPPA. In both algorithms, the convolution kernel parameters are optimized to represent a mapping function from undersampled zero-filled
k-space data $\mathbf{k}_0$ to reconstructed $\mathbf{k}$, where the missing lines are calculated from local correlations to neighboring multichannel k-space points. A similar approach is pursued by Wang et al.~\cite{Wang.2020}, who use an unrolled Convolutional Neural Network (CNN) with data-consistency layers to map from $\mathbf{k}_0$ to $\mathbf{k}$.

Here, we focus on a different reconstruction strategy, coined Deep Image Prior (DIP), which originates from the field of image restoration~\cite{Ulyanov.2018}. Similar to MR-reconstruction,
image restoration tasks can be formulated as an optimization problem. 
Instead of using a gradient descent method to search for an optimal solution $\mathbf{x}$ directly in the image space, DIP shifts the search to the parameter space of the Neural Network. 
The network weights are optimized, until the network is able to reconstruct the final image out of a fixed random noise input sample.
Ulyanov et al. showed, that a U-Net \cite{Ronneberger.2015} architecture is biased to prefer the reconstruction of natural images over less structured data. 
Therefore, when training the network to reconstruct a corrupted image $\mathbf{x}_0$, starting from randomly initialized parameters, it can be observed that the solution
$\mathbf{x}$ is reached with lower impedance than $\mathbf{x}_0$ during the training \cite{Ulyanov.2018}. 
The architecture serves as a statistical prior of $\mathbf{x}$.
DIP has shown competitive performance in image restoration tasks such as inpainting, denoising and super-resolution.
In MR reconstruction this is leveraged for non-linear interpolation of missing k-space data.
While RAKI only works for regular undersampling patterns, statistical patterns can achieve higher acceleration rates, as known from CS, and they can be used, e.g., for retrospective triggering in cine MRI.

Previous studies investigated this approach in combination with
learning from reference data to initialize the network weights from training on similar images~\cite{DiZhao.2021,Shen.2022,Gungor.2023} or by adding a trained discriminator network to the loss function~\cite{Korkmaz.2022, Elmas.2023}.
These approaches bear the risk of introducing bias from the samples used for this pre-training, which is why we focused on a method that can achieve good reconstruction results without the need of any previous training-data.

A training-free approach was also pursued in other studies, that focused on reconstruction of radial or spiral acquisition of dynamic (2D+t) MRI data~\cite{Yoo.2021,Hamilton.2023}. These data exhibit high temporal redundancy, which typically facilitate high acceleration rates. In this study, we demonstrate that the DIP approach can also achieve high acceleration even when dealing with less redundant input data, such as 2D Cartesian sampling.

Furthermore, improving the reconstruction by a combination of DIP with TV regularization of the loss-function was studied in non-MRI contexts like image denoising or CT reconstruction~\cite{Liu.2019,Baguer.2020,Liu.2023} or in quantitative MRI \cite{Slavkova.2023}.
The system matrix in these settings is very different to the image acquisition investigated here, which makes comparisons difficult and therefore dedicated investigations are necessary.
Darestani et al. investigated variations of the network architecture and their implications on reconstruction performance and presented \mbox{Conv}Decoder for reconstruction of 2D MRI with Cartesian undersampling~\cite{Darestani.2021}.

Here, we utilize the original U-Net architecture suggested for DIP \cite{Ulyanov.2018} and investigate, how different sparsity regularization terms in the loss function improve reconstruction results.
In the following, we illustrate the resulting SCAMPI architecture and demonstrate that our approach outperforms the DIP-reconstruction with a U-Net architecture and state-of-the art parallel imaging MRI reconstruction methods. Furthermore, it can also be applied in single channel reconstruction.
In comparison to ConvDecoder, it shows improved reconstruction quality with much fewer iterations.

\section{Methods}\label{sec:methods}

\subsection{Deep Image Prior}
Undersampling reconstruction can be illustrated as filling the missing data points of the undersampled measurement
\begin{equation}
	\mathbf{k}_0 = \mathcal{P} \, \mathbf{k} . 
\end{equation}
Here, ${ \mathbf{k} \in \mathbb{C}^{n_c \times N_x \times N_y}}$ is the full set of spatial frequency data.
$n_c$ is the number of coil channels and $N_{x, y}$ the number of k-space coefficients in readout and phase encoding direction. 
$\mathcal{P}$ is the undersampling projection operator, effectively setting certain lines in phase-encoding direction of $\mathbf{k}$ to zero. 

DIP on the other hand has shown promising results in various image reconstruction tasks.
However, convolutional layers, that are used in DIP, favor natural images by imposing local spatial correlations on different length scales.
But the k-space results from an integral transform and does not show these local structures.
Hence, the reconstruction is not searched directly in k-space by our approach.
Instead, the U-Net \cite{Ronneberger.2015} $f$ operates in image space.
By minimizing a loss-function, the parameter space of network weights~$\theta$ is searched, such that an image
${ \mathbf{x}=f_\theta(\mathbf{z}) }$
is built from a fixed but randomly initialized
input ${\textbf{z} \in \mathbb{R}^{2 \times N_x \times N_y}}$.
To iteratively find the optimal k-space reconstruction
${\hat{\mathbf{k}} = \mathcal{FS}\hat{\mathbf{x}} = \mathcal{FS}f_{\hat{\theta}}(\mathbf{z})}$, 
the parameters $\theta$ of the network $f$ are optimized, such that
\begin{equation} \label{eq:argmin}
	\hat{\theta} = \mathrm{arg}\,\underset{\theta}{\mathrm{min}} \; \mathcal{L}( f_\theta(\mathbf{z}), \mathbf{k}_0) \; ,
\end{equation}
where $\mathcal{L}$ is a cost function, typically the mean squared error. 
$\mathbf{k}$ is then obtained from $\mathbf{x}$ as the coil sensitivity-weighted Fourier transform
${ \mathbf{k} = \mathcal{FS}\mathbf{x}}$, 
where $\mathcal{F}$ is the Fourier transform and 
$\mathcal{S}$ represents the coil sensitivity profiles. 

\subsection{The SCAMPI network architecture and setup}

For SCAMPI we used a U-Net architecture like other DIP approaches (s.~\figref{fig:architecture_and_loss}~A). Details on the layer parameters are provided in Supplementary Materials Tab.~1.

In a series of pre-studies, we investigated, how the cost-function $\mathcal{L}$ can be adapted to improve reconstruction performance. Our findings and the deduced loss-function are elaborated in \secref{sec:methods_costfunc}.
Fourier transforming the network output before passing it to $\mathcal{L}$ allows to implement a loss functional that compares $\mathbf{k}$ against the measured $\mathbf{k}_0$, while using the network's convolutions in the image domain as an appropriate prior for the image statistics and spatial correlations. 

%
After stopping the iteration, the final multicoil reconstruction is given by $\hat{\mathbf{k}}$ after ensuring data-consistency (DC) with the sampled k-space points $\mathbf{k}_0$ (s.~\secref{sec:methods}).
The DC function in \eqref{eq:cost_func} is formalized for each data point $i$ by
\begin{equation} \label{eq:DC}
	\left(\mathrm{DC}(\hat{\mathbf{k}}, \mathbf{k}_0;\mathcal{P})\right)_i =
	\left\{\begin{aligned}
		&(\mathbf{k}_0)_i \quad &&\mathrm{if}\; (\mathcal{P})_i = 1\\
		&(\hat{\mathbf{k}})_i \quad &&\mathrm{else.}
	\end{aligned}\right.
\end{equation}
Finally, the image space reconstruction is obtained by phase-sensitive combination of the Fourier transform of the multicoil k-space data, as described by Roemer et al.~\cite{Roemer.1990}.
\begin{algorithm}[t]
	\caption{k-Space reconstruction with SCAMPI.}\label{alg1}
	\begin{algorithmic}
		\Require Undersampled k-space ${\mathbf{k}_0 \in \mathbb{C}^{n_c \times N_x \times N_y}}$, number of epochs $n$, 
		learning rate $l$
		\State Determine sampling pattern $\mathcal{P}$ from non-zero coefficients in ${\mathbf{k}_0}$
		\State Estimate coil sensitivity profiles $\mathcal{S}$ from $\mathbf{k}_0$ with \mbox{ESPIRiT}
		\State Randomly initialize ${\mathbf{z} \in \mathbb{R}^{2 \times N_x \times N_y}}$ and 
		$\theta_{i=0}$
		\For{$n$ iterations}
		\State $\mathbf{x}_{i} = f_{\theta_i}(\mathbf{z}) $ with $\mathbf{x} \in \mathbb{C}^{N_x \times N_y} $
		\State $\theta_{i+1} \leftarrow \operatorname{ADAM}\left(\theta_i,\, \mathcal{L}(\mathbf{x}_i, \mathbf{k}_0),\, l\right) $
		\EndFor
		\State $\hat{\mathbf{x}} = f_{\theta_{n}}(\mathbf{z})$
		\State $\hat{\mathbf{k}} = \mathrm{DC}(\mathcal{FS}\hat{\mathbf{x}}, \mathbf{k}_0;\,\mathcal{P})$
		\Ensure Final reconstruction $\hat{\mathbf{k}} \in \mathbb{C}^{n_c \times N_x \times N_y}$
	\end{algorithmic}
\end{algorithm}
\begin{figure*}[btp]
	\centering
	\includegraphics[width=.71\linewidth]{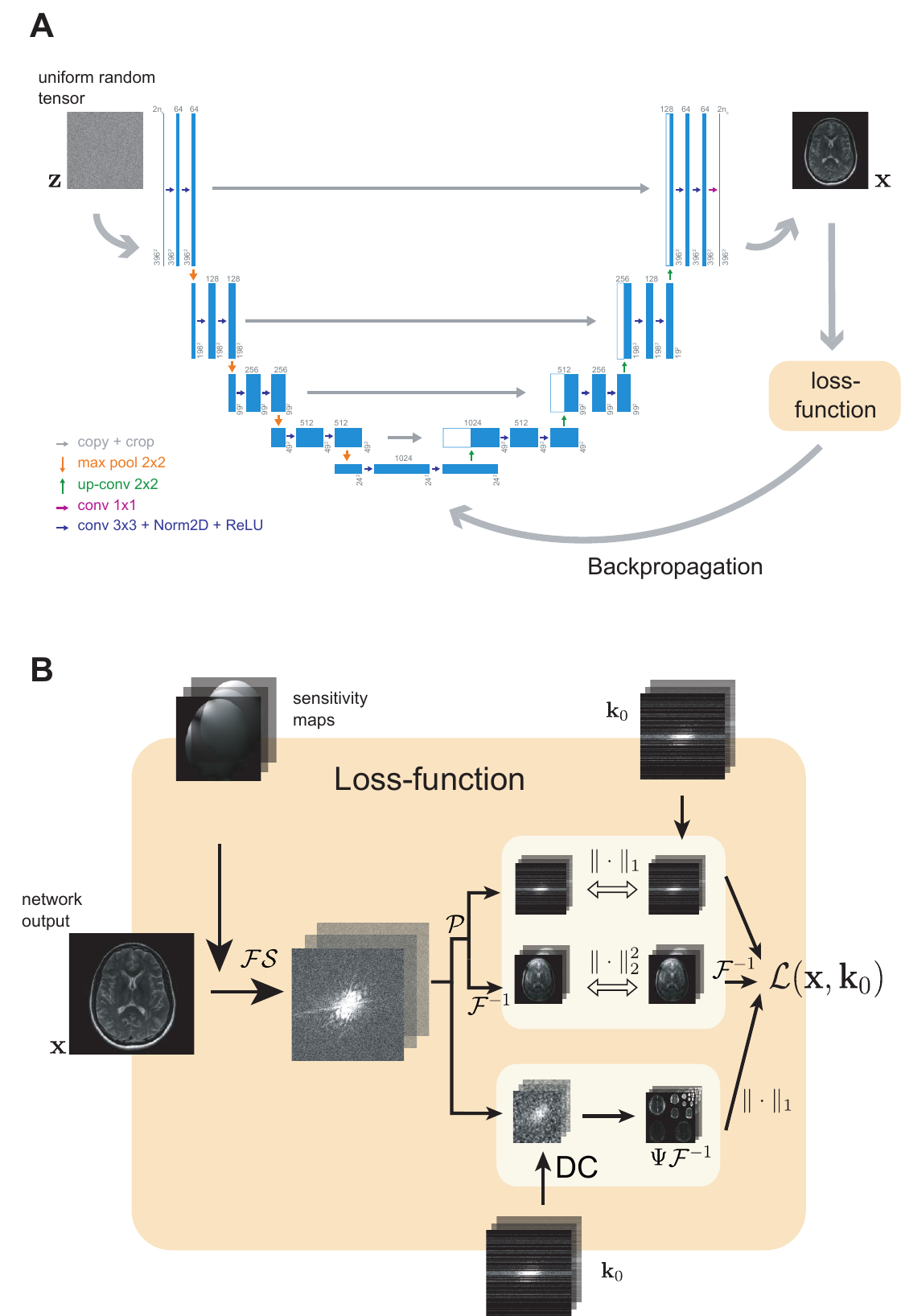}
	\caption{
		The SCAMPI architecture uses
		\textbf{A} a U-Net with 2D convolutions conv~3x3, where 3x3 describes the 2D convolutional kernel's size. The number of filter maps is depicted above the layers and reported in Supplementary Materials, Tab.~1.
		2D normalization layers and Rectified Linear Units(ReLU) are placed between the convolutional layers.
		Maximum pooling (max~pool~2x2) with kernel size 2x2 is used for down-sampling. Bilinear interpolation is used for up-sampling (up-conv~2x2). Zero-padding is used in convolutional layers to preserve the input images' sizes. 
		U-Net's final output layer returns a $\mathbb{R}^{2\times N_x \times N_y}$ tensor.
		The first dimension's two channels are treated as real and imaginary part of the image 
		$\mathbf{x} \in \mathbb{R}^{N_x \times N_y}$.
		\textbf{B}
		In the loss function $\mathcal{L}$, the transformed network output ${ \mathbf{k}=\mathcal{FS}\mathbf{x} }$ is compared to $\mathbf{k}_0$ in the measured data-points points and subjected to a sparsity constraint (here, L1 norm of wavelet coefficients is shown as an example).
	}
	\label{fig:architecture_and_loss}
\end{figure*}
\subsubsection{Loss function}\label{sec:methods_costfunc} 
\label{sec:cost_func}
%
%

During the training process as given by \eqref{eq:argmin}, the network parameters~$\theta$ are tuned by gradient backpropagation from the loss-function~$\mathcal{L}$.
To meet the requirements for the non-linear k-space interpolation, we heuristically assembled a dedicated loss function. 
Each term has a specific impact on the overall reconstruction result.
The L1 norm in k-space is more resilient to outliers and causes errors in the k-space center to be less significant.
Furthermore, we incorporated a L2 loss term in image domain to enforce appearance-consistency \cite{Zhou.2022}.
For additional sparsity regularization, we implemented Total Variation (TV), as well as L1 norm of the wavelet coefficients (L1W). 

\begin{figure*}[btp]
	\centering
	\includegraphics[width=.90\linewidth]{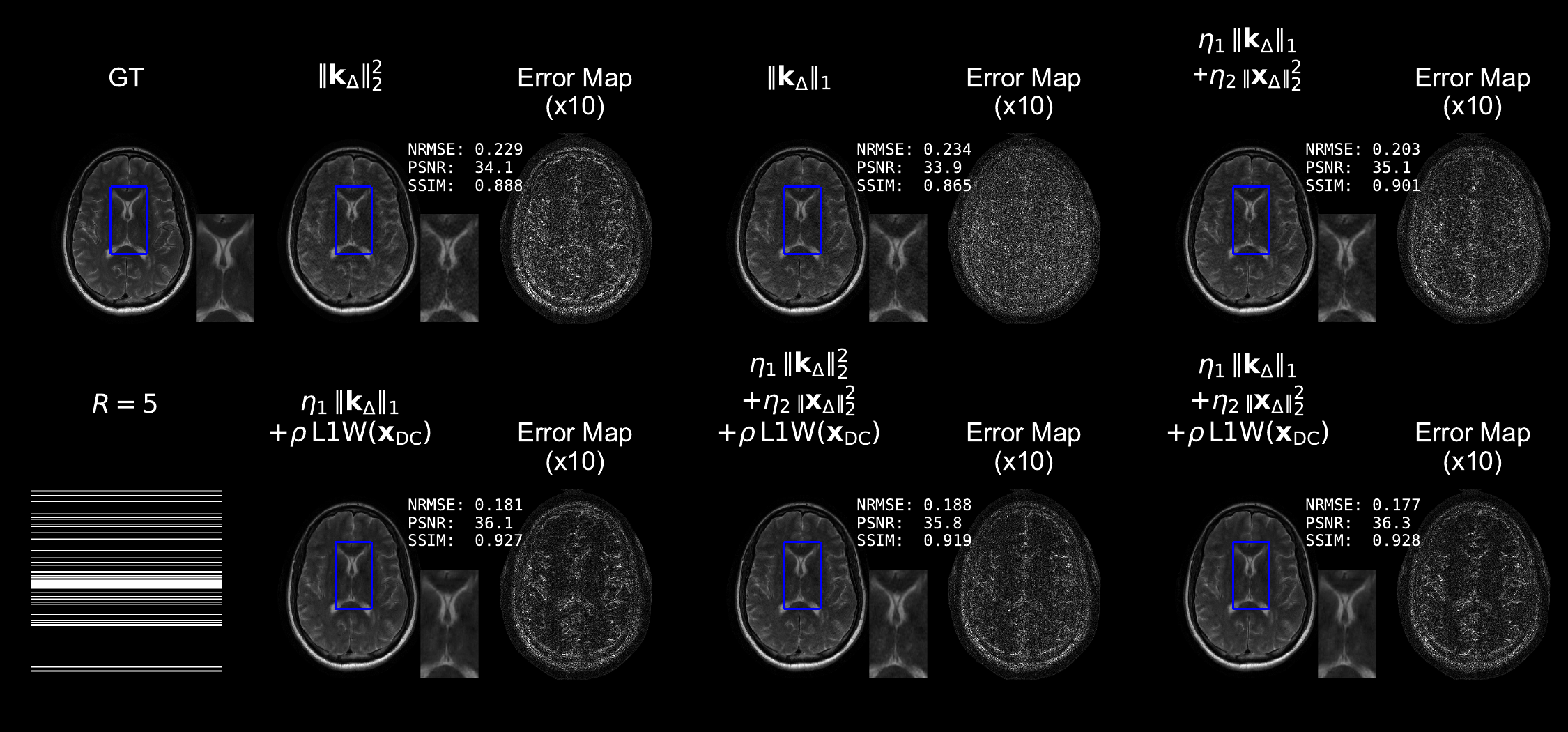}
	\caption{
		Multi-coil ($n_c=12$) reconstruction of undersampled k-space data (${R=5}$) with U-Net and different loss terms.
		The shorthand notations denote:
		${\mathbf{x}_\Delta = \mathcal{PFS}\mathbf{x} - \mathbf{k}_0}$,
		${\mathbf{k}_\Delta = \mathcal{F}^{-1}\left(\mathcal{PFS}\mathbf{x} - \mathbf{k}_0 \right)}$, 
		${\mathbf{x}_\mathrm{DC} = \mathcal{F}^{-1}\mathrm{DC}(\mathcal{FS}\mathbf{x}, \mathbf{k}_0)}$.
		The weighting parameters $\eta_1, \eta_2, \rho$ were tuned heuristically. 
		Error Map shows relative difference ($\times 10$) to the Ground Truth (GT) reconstruction. 
		The undersampling pattern in k-space is illustrated
		below the GT image. 
		Quantitative error measures NRMSE, PSNR and SSIM, compared to the GT are reported next to each reconstruction image. 
		Inlay image shows zoom-in of the periventricular region (blue frame).\\
	}
	\label{fig:loss_term_figure}
\end{figure*}

We evaluated different combinations of loss terms for optimized reconstruction of an undersampled ($R=5$) multicoil ($n_c=12$) image from the FastMRI brain dataset~\cite{Zbontar.21.11.2018} (s.~\secref{sec:methods} for details).
Reconstruction quality was assessed visually and quantitatively, using NRMSE, PSNR and SSIM (as 
introduced in \secref{sec:image_stats_loss_functionals}),
which were calculated from magnitude images of the reconstructions against the ground truth (GT) reference image.
A selection of results is depicted in \figref{fig:loss_term_figure}. 
Ascertaining data-fidelity with L1 leads to noisier images, than with L2 penalty. 
However, when combining multiple loss terms, the combination of data-fidelity with L1, appearance-fidelity in image space with L2 and sparsity regularization achieves the best results. 

Based on these results, we proceed with the combination of loss terms, that is depicted in \figref{fig:architecture_and_loss}~B and is given as
\begin{equation}\label{eq:cost_func}
	\begin{aligned}
		\mathcal{L}(\mathbf{x}, \mathbf{k}_0) =\, & \eta_1 \,  \left\Vert \mathcal{PFS}\mathbf{x} - \mathbf{k}_0\right\Vert_1 \\
		+& \eta_2 \,  \left\Vert \mathcal{F}^{-1}\left(\mathcal{PFS}\mathbf{x} - \mathbf{k}_0 \right) \right\Vert_2^2\\
		+& \rho \,\mathcal{R}\left[
		\mathcal{F}^{-1}\mathrm{DC}(\mathcal{FS}\mathbf{x}, \mathbf{k}_0)
		\right] \; .
	\end{aligned} 
\end{equation}

The first two terms of $\mathcal{L}$ assert the reconstruction's data and appearance fidelity to $\mathbf{k}_0$.
The last term enforces transform-domain sparsity of $\mathbf{x}$.
The DC function in the last term proved necessary to avoid divergence of the loss during the first iterations of training, since the random input is not sparse in the transform domain.
Settings for the weighting parameters $\eta_{1,2}$ and $\rho$ are reported in \secref{sec:methods}.

\subsubsection{Calibration-free reconstruction}
\label{sec:methods_calibfree}
For multicoil data with $n_c$ coils, the number of output filter maps of the network can be adjusted to output
${  \mathbf{X}  \in \mathbb{C}^{n_c \times N_x \times N_y} }$
instead of 
${ \mathbf{x} \in \mathbb{C}^{N_x \times N_y} }$.
This allows to build an algorithm that does not require previously calibrated coil sensitivity profiles, but inherently outputs a multicoil k-space approximation
${ \mathbf{k} = \mathcal{F} \mathbf{X} }$. Then, ${ \mathbf{x} \in \mathbb{C}^{N_x \times N_y} }$ 
is given as the root sum of squares of $\mathbf{X}$. 

\subsection{Implementation of algorithms}\label{sec:implementation-of-algorithms}
%
Python~3 was used for all algorithms and simulations described in this work.
The SCAMPI network and the training pipeline were built using the PyTorch library \cite{AdamPaszke.2017}. 
For optimization of the network weights, the ADAM~\cite{Kingma.22.12.2014} algorithm with a learning rate $l$ is used.
The developed reconstruction algorithm is summarized in Algorithm~\ref{alg1}.
The U-Net implementation builds on publicly available code \cite{milesial.08.08.2022}. Details on the layer parameters can be found in Supplementary Materials, Tab.~1.
All complex-valued tensors were implemented inside the network as tensors in $\mathbb{R}^2$, representing real and imaginary part of the data.
If the output-tensor included coil data (in the calibration-free case), the complex and coil channels were stored in a shared channel inside the tensor, i.e., a multicoil image with $n_c$ coils is represented as a $2\, n_c$ channel tensor.
The coil sensitivity profiles $\mathcal{S}$, that are required for some of the reconstructions were obtained via ESPIRiT \cite{Uecker.2014} calibration, which was implemented in BART~\cite{MartinUecker.2018}.

SCAMPI reconstructions were generated by $1000$ iterative steps of updating the network weights.
The learning rate $l$ as well as the weighting parameters of the loss function in \eqref{eq:cost_func}, $\eta_{1,2}$ and $\rho$ were tuned heuristically and are reported in \tabref{tab:hyperparams}.
To determine the required number of iterations, the training loss of the reconstruction against the undersampled measurement data was recorded during training. $1000$ epochs were determined as a good trade-off between reconstruction time and quality for all settings (see Suppl., Sec.~S1.3 for details).

For all reported comparisons we used the PICS algorithm implemented in SigPy~\cite{sigpy.23.01.2022}, which optimizes
\begin{equation}
	\min_\mathbf{x} \frac{1}{2} \left\Vert \mathcal{PFS}\mathbf{x} - \mathbf{k}_0\right\Vert_2^2 + \lambda \; \mathrm{L1W}[\mathbf{x}]
\end{equation}
via iterative soft thresholding.
We iterated for 
$500$ steps and heuristically tuned weighting parameter $\lambda=0.5$ of the sparsity regularization term L1W.

For TV, we used a finite gradient operator with a one-pixel-shift in both spatial dimensions. For L1W, we used a wavelet transformation $\Psi$ with Daubechies-Wavelets (D4) as basis functions and decomposition level of $5$.

For ENLIVE, we chose the parameters ($a=240, b=40, m=1$), as introduced by \cite{Holme.2019}. We chose the number of iterations $i=9$, because for more iterations we observed an increase of high frequency noise, as described by the authors.

For ConvDec, we used the settings, given in the code example by the authors for the knee images: $160$ channels, an input size of ($8$, $4$) and a kernel size of $3$~\cite{Darestani.2021}.
For the brain images, we set the input size to
($4$, $4$), due to the smaller and quadratic matrix size of the k-space data.
The network was optimized with ADAM optimizer and learning rate $l=0.008$ for $20000$ iterations.
To further investigate the observed differences between ConvDecoder, which employs a different network architecture, and SCAMPI, we also considered reconstructions with a DIP approach using the same U-Net architecture without sparsity-regularization by using MSE as loss function.

Reconstructions were performed on a
Linux system with Intel(R) Xeon(R) Silver 4214R CPU and Nvidia(R) Titan RTX GPU. 

\begin{table*}[btph]%
	\centering
	\caption{
		Chosen hyperparameters for SCAMPI reconstructions presented in \secref{sec:results}. Learning Rate $l$ and weighting factors
		$\eta_1$ for k-space, $\eta_2$ for image domain loss and $\rho$ for sparsity constraints Total Variation (TV) and L1 norm of Wavelets (L1W).
		\label{tab:hyperparams}
	}
	\begin{tabular*}{.78\textwidth}{@{\extracolsep\fill}|l|cccc|cc|@{\extracolsep\fill}}
		\toprule
		\toprule
		~ & ~ & ~ & TV & ~ & L1W & ~  \\
		\midrule
		Multicoil & $\eta_1$ & $\eta_2$ & $l$ & $\rho$ & $l$ & $\rho$ \\
		\midrule
		with coil sensitivities & 20 & 1 & 0.03 & $3\times 10^{-8}$ & 0.03 & $3\times 10^{-8}$  \\ 
		\midrule
		without coil sensitivities & 20 & 1 & 0.03 & $1\times 10^{-8}$ & 0.03 & $1\times 10^{-7}$ \\
		\bottomrule
		\toprule
		Single coil & 20 & 1 & 0.01 & $5\times 10^{-6}$ & 0.01 & $5\times 10^{-6}$ \\
		\bottomrule
	\end{tabular*}
\end{table*}

\subsection{Reconstruction data}\label{sec:reconstruction-data}
Data used in the preparation of this article were obtained from the NYU fastMRI initiative database (\url{fastmri.med.nyu.edu}) \cite{Knoll.2020.fastMRI, Zbontar.21.11.2018}.
We did not perform experiments on humans or human tissues.

\subsubsection{Multicoil data}
\label{sec:reconstruction-data_multicoil}
To deduce the hyperparameter settings for multicoil-reconstruction reported in Tab.~\ref{tab:hyperparams}, the reconstruction performance was evaluated on an arbitrarily chosen, axial T2-weighted image from the fastMRI brain dataset
(file name \texttt{file\textunderscore brain\textunderscore AXT2\textunderscore 209\textunderscore 6001331}).
One bottom slice from the fully sampled multi-slice dataset was taken as the ground truth (GT) and retrospectively undersampled.
To save memory, the field of view was cropped to $396^2$ pixels.
For Cartesian undersampling, we pseudo-randomly selected readout lines from a Gaussian probability density distribution with maximum in \mbox{k-space} center. The sampling algorithm was designed to assure that the central $18$ lines were always sampled. 
Multicoil (${n_c = 12}$) reconstruction performance was assessed for
${R=3, 5, 8}$. 
%
The robustness of our results was then investigated on a subset of $200$ different samples, selected randomly from the FastMRI brain dataset (validation set).
Different MRI contrast modalities (FLAIR, T1, T2) were present in the subset. The number of receiver coils varied between ${n_c=4}$ and ${n_c=20}$.
From each multi-slice dataset, the lowest slice was selected and cropped to shape $N^2$, where ${N=\min(N_x, N_y)}$ and $N_{x,y}$ is the number of pixels in the two spatial dimensions.
For each sample, a new undersampling pattern was randomly generated.
Reconstruction was performed as outlined for the single sample described above, and error metrics of SCAMPI and baseline reconstructions were compared. 
\subsubsection{Assessment of parallel imaging noise}
\label{sec:methods_gfactor}

To quantify the noise amplification of the different reconstruction methods, we employed the pseudo multiple replica method, which was originally suggested to quantify the g-factor of linear parallel imaging methods.~\cite{Robson.2008}. 
More recent studies did also use it to estimate noise in Neural Network based reconstructions~\cite{Kleineisel.2023}.
We did $n_{\text{MC}}=30$ pseudo repetitions of the same reconstruction and calculated the 
pixel-wise standard-deviation $\mathbf{\sigma}$ between the separate reconstructions. 
The estimated g-factor maps are then given by 
\begin{equation}
	\label{eq:g_factor}
	\mathbf{g} = \frac{ \mathbf{\sigma}_{\text{acc}}}{\mathbf{\sigma}_\text{N} \,\sqrt{R}} \quad ,
\end{equation}
where $\mathbf{\sigma}_\text{acc}$ denotes the point-wise standard deviation of the undersampled reconstruction (acceleration $R$) with additional Gaussian noise from a standard normal distribution added to the k-space and $\mathbf{\sigma}_\text{N}$ that of pure Gaussian noise alone.

\subsubsection{Single coil data}
\label{sec:methods_single_coil_data}
Furthermore, single coil (${n_c = 1}$) reconstruction for ${R=2, 3}$ was investigated
on a central slice from the FastMRI single coil knee validation dataset \cite{Zbontar.21.11.2018}
(\texttt{file1001221}).

A quantitative analysis of the reconstruction quality at ${R=2, 3}$ was performed on $200$ randomly selected images from the dataset, which were processed in the same manner as the multicoil data described above.

\subsection{Image error metrics} \label{sec:image_stats_loss_functionals}

For quantitative evaluation of our results we used the Normalized Root Mean Squared Error (NRMSE), Peak Signal Noise Ratio (PSNR) and Structural Similarity Index (SSIM), which are frequently used in the literature.

\section{Results}\label{sec:results}
%

\subsection{Parallel Imaging Reconstructions}

In the following, we visualize the reconstruction results on our subset from the FastMRI brain dataset~\cite{Zbontar.21.11.2018}, as described in \secref{sec:reconstruction-data_multicoil}. 

\begin{figure}[tbh]
	\centering
	\includegraphics[width=0.98\columnwidth]{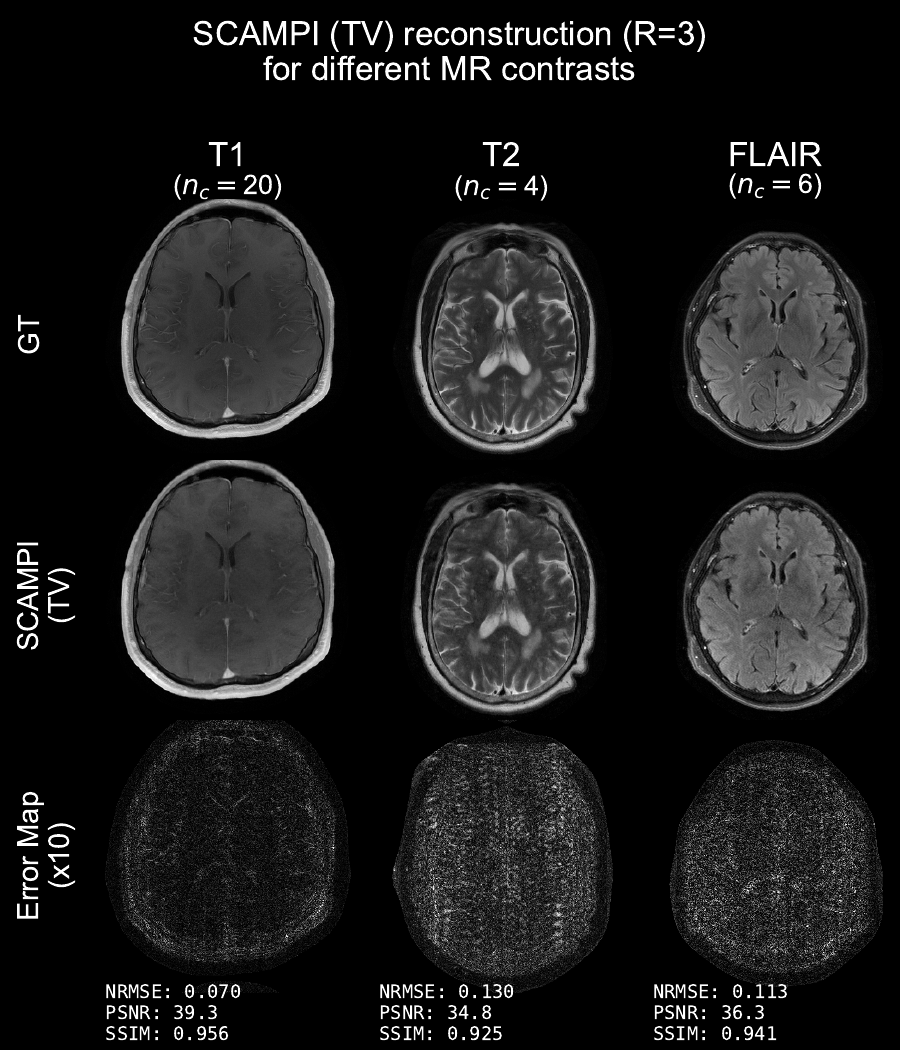}
	\caption{
		Scampi (TV regularization) reconstructions for acceleration rate $R=3$ of samples with different imaging contrasts and number of receiver coils $n_c$. 
		Quantitative error measures NRMSE, PSNR and SSIM, compared to the GT are reported below the difference map (scaled $\times10$).
		\label{fig:contrasts}
	}
\end{figure}

Fig.~\ref{fig:contrasts} illustrates that our setup is capable of reconstructing different imaging contrasts T1, T2 and FLAIR, present in the dataset. 
We compared the SCAMPI approach that uses ESPIRiT-estimated coil sensitivity profiles against PICS, which also requires sensitivity map estimates.

\begin{figure*}[btp]
	\centering
	\begin{subfigure}{.98\textwidth}
		\centering
		\includegraphics[width=1\linewidth]{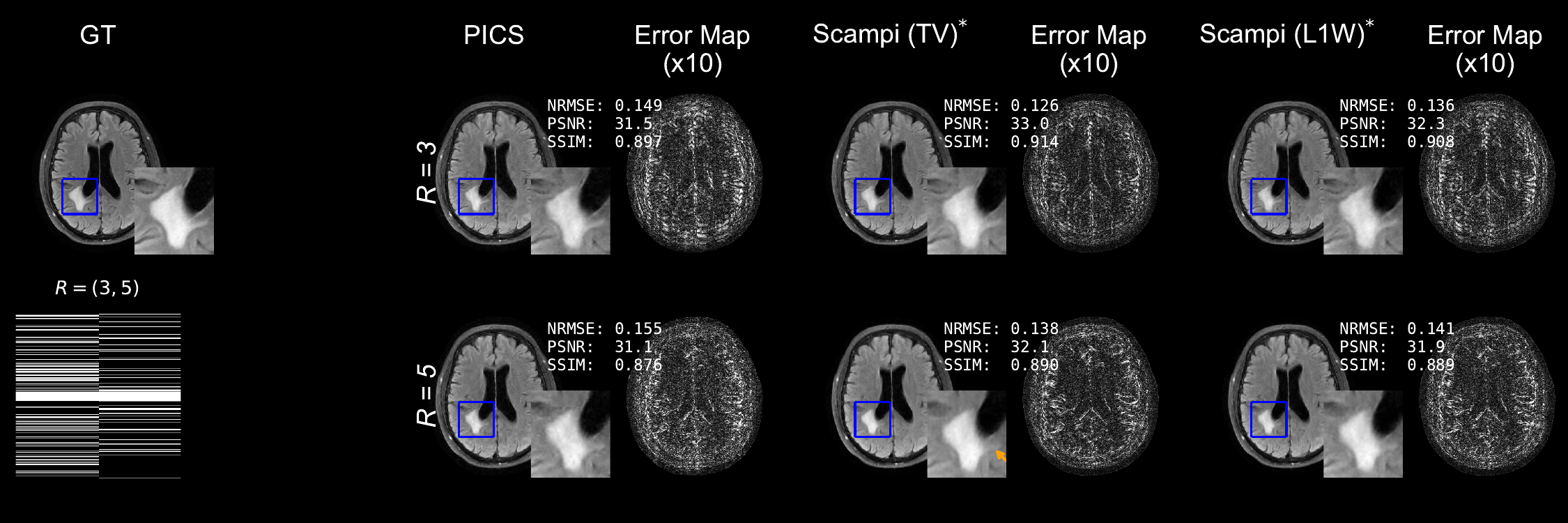}
		\label{fig:pics_v_scampi}
	\end{subfigure}%
	\\
	\begin{subfigure}{.98\textwidth}
		\centering
		\includegraphics[width=1\linewidth]{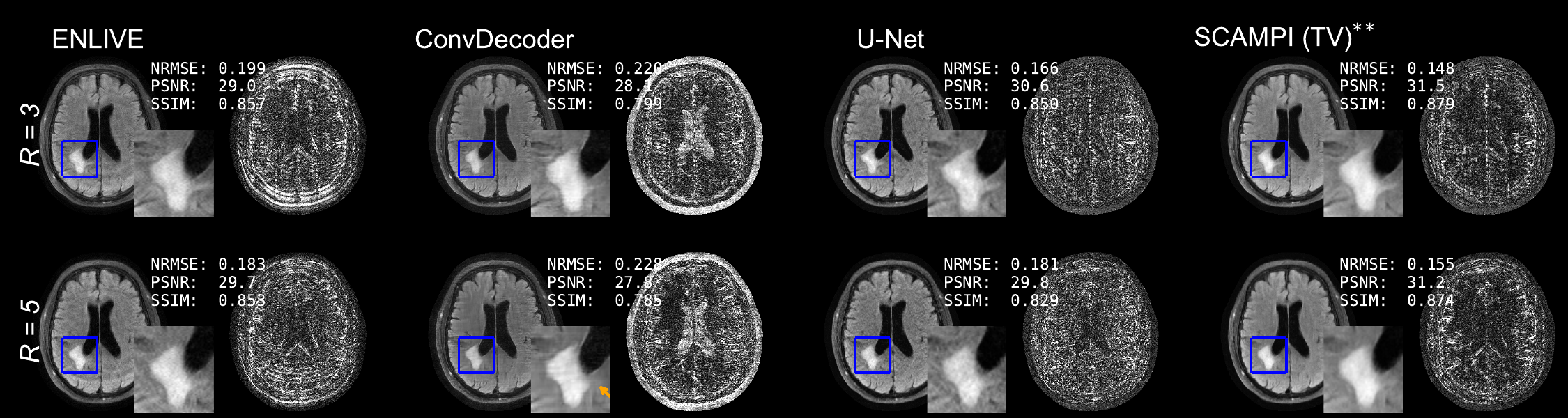}
		\label{fig:compare_ohne_spulen}
	\end{subfigure}
	\caption{
		Comparison of FLAIR image reconstruction at different acceleration rates $R$. The reconstructions at ${R=8}$ are not displayed in the section above, because all of the methods failed to produce clinically acceptable reconstructions.
		Error Map shows relative difference ($\times 10$) to the Ground Truth (GT). The undersampling pattern in k-space for different $R$ is illustrated below the GT image. 
		Quantitative error measures NRMSE, PSNR and SSIM, compared to the GT are reported next to each reconstruction. 
		The \textbf{upper panel} shows reconstruction methods with coil-sensitivity maps: PICS (L1W regularization) and SCAMPI (TV or L1W, $1000$ iterations).
		The \textbf{lower panel} displays the root sum of squares of calibration-free reconstruction with ENLIVE, ConvDecoder ($20000$ iterations), U-Net ($1000$ iterations) and SCAMPI ($1000$ iterations)
		with Total Variation (TV) sparsity regularization.
		No coil sensitivity estimates are used for these reconstructions.
		The orange arrow indicates, how noise amplification in the reconstructed images can lead to deteriorated fine structure resolution.		
		\label{fig:multicoil_recos}
	}
\end{figure*}

Details on parameters are reported in \secref{sec:methods}.
PICS showed better results with L1W regularization, which is compared against SCAMPI with TV or L1W regularization in \figref{fig:multicoil_recos}.

For both regularization terms, TV and L1W, SCAMPI shows less residual error in the error map and better image quality metrics than the PICS reconstruction.

As noted in \secref{sec:methods_calibfree}, the SCAMPI method can be adapted to allow a calibration-free reconstruction, i.e., without requiring ESPIRiT-estimated coil sensitivity profiles.
In this case, magnitude images were obtained from the multicoil images by calculating the root sum of squares. 

As a baseline, we compare against other calibration-less reconstruction methods: the low-rank based method ENLIVE \cite{Holme.2019} and ConvDecoder~\cite{Darestani.2021}, another DIP-based approach, whose architecture deviates from the original one and that does not use sparsity-regularization. Furthermore, we studied the effect of our refined loss-term by comparing against a DIP with standard U-Net architecture and MSE penalty. 
Details on (hyper)parameter settings are described in~\secref{sec:implementation-of-algorithms}.
To filter random noise in the low signal-region outside the subject's anatomy, a signal threshold was applied to the images.

Reconstruction results are displayed in
\figref{fig:multicoil_recos}.
As can be seen from the residual error maps and the image metrics, calibration-free U-Net can achieve better reconstruction results than the baseline methods ENLIVE and \mbox{Conv}Decoder. 
Compared to the U-Net with MSE penalty, the extended loss function of SCAMPI leads to noticeably better reconstruction results. 

Note that for these results, SCAMPI is iterated 20-times fewer than ConvDecoder, resulting in higher reconstruction speed of ca.~$98\,\mathrm{s}$ versus ca.~$21\,\mathrm{min}$ (cf.~\tabref{table:reco_times}).
A comparison with the ESPIRiT-calibrated methods shows, that the best reconstruction results with less residual artifacts can be achieved with the SCAMPI algorithm using coil sensitivity information.

\subsubsection{Assessment of reconstruction noise}
\label{sec:results_multicoil_noise}

While noise amplification in parallel imaging methods is well-described, the behavior of neural networks is less studied. In our results (s. previous section), the ConvDecoder reconstructions appear noisier than the U-Net and SCAMPI reconstructions, which could explain the inferior image metrics even though \mbox{Conv}Decoder does show an effective undersampling artifact suppression. 
Nonetheless, image noise can relevantly impede fine structure resolution, as is demonstrated by the fine tail of the hyperintense periventricular lesion in the image inlay (orange arrow in \figref{fig:multicoil_recos}).

\begin{figure}[tbh]
	\centering
	\includegraphics[width=0.98\columnwidth]{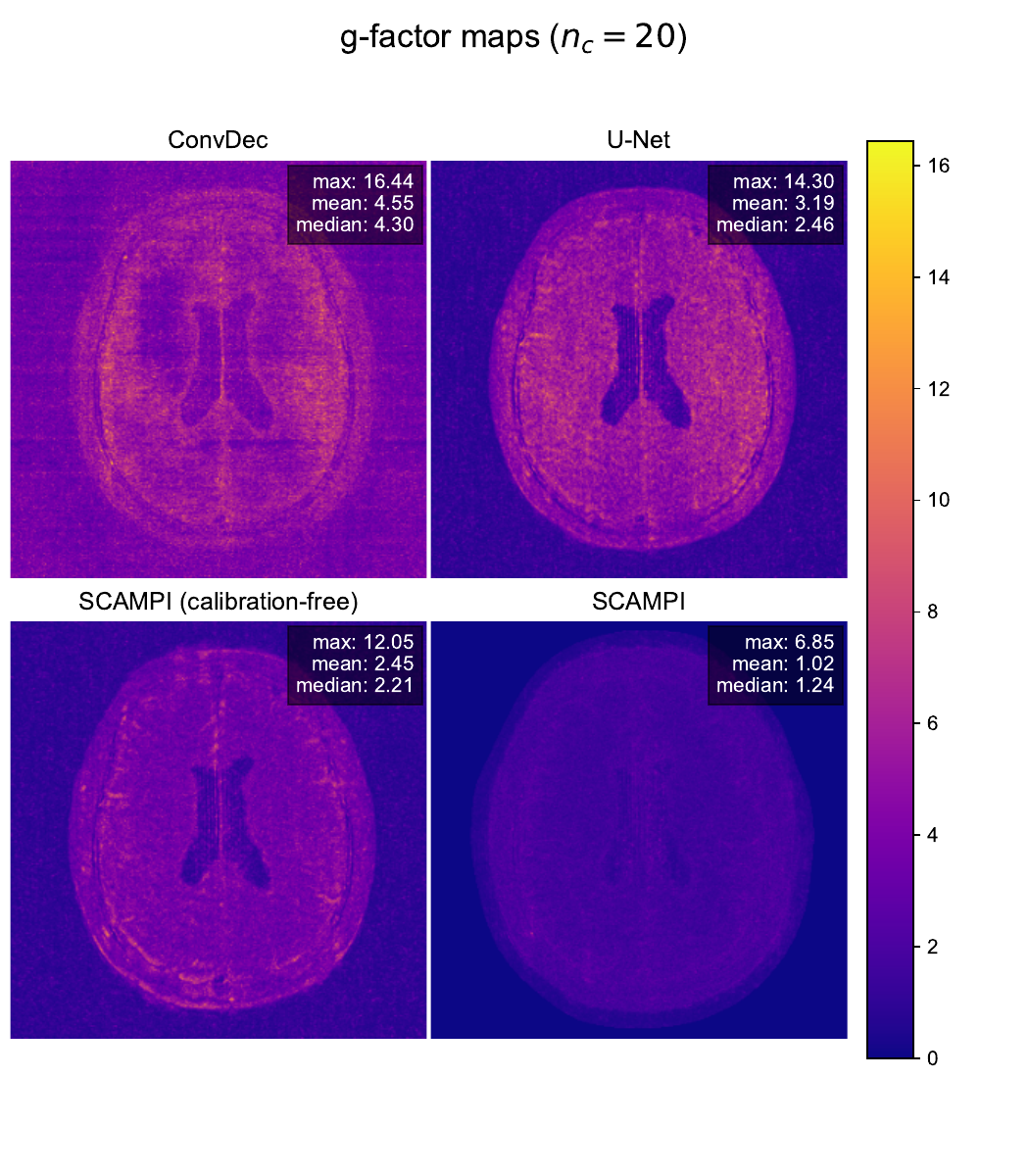}
	\caption{
		Reconstruction noise assessment for different neural network methods at ${R=3}$ as described by \eqref{eq:g_factor} in \secref{sec:methods_gfactor}. There is substantial noise in the \mbox{Conv}Decoder and U-Net reconstructions, while it is lower with SCAMPI, especially the calibrated one. 
		\label{fig:gfactor}
	}
\end{figure}

We quantified the reconstruction noise as described in \secref{sec:methods_gfactor} at ${R=3}$ for the different neural network methods. 
The results in \figref{fig:gfactor} confirm our findings reported in the previous section, where noise was especially prominent in the ConvDecoder reconstruction.

\begin{table}[tb]
	\centering
	\caption{
		Reconstruction time $t_\text{r}$ for different reconstruction algorithms and numbers of reconstructed coils $n_c$ of the $396^2$ pixel brain image. Estimated by averaging over 5 reconstructions.\\
		$\dagger$: without GPU acceleration.
		\label{table:reco_times}
	}
	\begin{tabular}{|c| l   |c|}
		\midrule
		$n_c$ & Algorithm & $t_\text{r}$ \\
		\midrule
		1 & ENLIVE & 13 s$^\dagger$ \\ 
		~ & CS (TV) & 1.9 s \\
		~ & CS (L1W) & 8.1 s\\
		~ & ConvDecoder & 16 min\\ 
		~ & U-Net & 90 s \\
		~ & SCAMPI (TV) & 90 s \\
		~ & SCAMPI (L1W) \mbox{      } & 116 s \\
		\midrule
		12 & ENLIVE & 23 s$^\dagger$ \\
		~ & PICS (TV) & 1.9 s\\
		~ & PICS (L1W) & 8.5 s\\
		~ & ConvDecoder & 22 min  \\
		~ & U-Net & 98 s \\
		~ & SCAMPI (TV) & 98 s \\
		~ & SCAMPI (L1W) & 258 s \\
		\bottomrule
	\end{tabular}	
\end{table}

\subsubsection{Quantitative analysis}\label{sec:multicoil_quantitative-analysis}

A quantitative analysis was performed on a subset of $200$ samples from the FastMRI dataset, as described in \secref{sec:reconstruction-data}.
Reconstruction was performed for PICS, ENLIVE, \mbox{Conv}Decoder, U-Net and SCAMPI (with and without calibration).
Details on reconstruction parameters are described in \secref{sec:implementation-of-algorithms}.
As can be seen from \tabref{tab:stats_multicoil}, the best image metrics are achieved with the calibrated SCAMPI reconstruction.
The calibration-free SCAMPI network also outperforms the reference method \mbox{Conv}Decoder, except in NRMSE and PSNR at ${R=8}$. 

\begin{table*}
	\centering
	\caption{
		Mean image quality metrics for PICS (L1W regularization), ENLIVE, \mbox{Conv}Decoder and SCAMPI on a random subset of $200$ samples from the FastMRI multicoil brain dataset. Bold values highlight the method with best metrics for each $R$. Asterisks indicate significance level of differences between the respective method and the best one, 
		determined by a Mann-Whitney U rank test: $^{-}: p > 0.4$, $^{\ast}: p < 0.05$, $^{\ast\ast}: p < 0.001$.
		Error margins denote $95\%$ confidence interval. \\
		$^\text{a}$: calibration-free. $^\text{b}$: coil-calibrated
		\label{tab:stats_multicoil}
	}
	\begin{tabular*}{.90\textwidth}{@{\extracolsep\fill}|l|l|ccc|@{\extracolsep\fill}}
		\toprule
		\toprule
		R & Method & NRMSE & PSNR & SSIM \\ 
		\toprule
		3 & PICS 							 & 0.169 ± 0.016 $^{\ast\ast}$	     & 34.83 ± 0.51 $^{\ast\ast}$ & 0.8894 ± 0.0088 $^{\ast}$\\ 
		~ & ENLIVE 							 & 0.2159 ± 0.0070 $^{\ast\ast}$	 & 32.22 ± 0.36 $^{\ast\ast}$ & 0.8641 ± 0.0074 $^{\ast}$\\ 
		~ & ConvDecoder 					 & 0.1760 ± 0.0080 $^{\ast\ast}$	 & 34.11 ± 0.43 $^{\ast\ast}$ & 0.8672 ± 0.0078 $^{\ast\ast}$\\ 
		~ & U-Net 							 & 0.1849 ± 0.0050 $^{\ast\ast}$     & 33.49 ± 0.36 $^{\ast\ast}$ & 0.8688 ± 0.0070 $^{\ast\ast}$\\
		~ & SCAMPI TV$^\text{a}$				 & 0.1653 ± 0.0046 $^{\ast\ast}$	 & 34.47 ± 0.36 $^{\ast\ast}$ & 0.8911 ± 0.0058 $^{\ast\ast}$\\ 
		~ & SCAMPI L1W$^\text{a}$			 & 0.1707 ± 0.0048 $^{\ast\ast}$	 & 34.19 ± 0.36 $^{\ast\ast}$ & 0.8840 ± 0.0061 $^{\ast\ast}$\\ 
		~ & SCAMPI TV$^\text{b}$		 & \textbf{0.1425 ± 0.0063}	 & \textbf{35.96 ± 0.45} & \textbf{0.9181 ± 0.0065} \\ 
		~ & SCAMPI L1W$^\text{b}$			 & 0.1485 ± 0.0067 $^{-}$               & 35.62 ± 0.45 $^{-}$           & 0.9123 ± 0.0069 $^{-}$\\
		\midrule
		5 & PICS         					 & 0.271 ± 0.018 $^{\ast\ast}$	& 30.53 ± 0.44 $^{\ast\ast}$ & 0.8291 ± 0.0097 $^{\ast\ast}$ \\ 
		~ & ENLIVE 							 & 0.331 ± 0.010 $^{\ast\ast}$	& 28.47 ± 0.29 $^{\ast\ast}$ & 0.815 ± 0.0068 $^{\ast\ast}$ \\ 
		~ & ConvDecoder						 & 0.2378 ± 0.0090 $^{\ast\ast}$	& 31.46 ± 0.40 $^{\ast}$ & 0.8312 ± 0.0090 $^{\ast\ast}$ \\
		~ & U-Net					 		 & 0.2983 ± 0.0082$^{\ast\ast}$	& 29.34 ± 0.33 $^{\ast\ast}$ & 0.7995 ± 0.0084 $^{\ast\ast}$ \\  
		~ & SCAMPI TV$^\text{a}$				 & 0.2560 ± 0.0077 $^{\ast\ast}$	& 30.71 ± 0.34 $^{\ast\ast}$ & 0.8489 ± 0.0073 $^{\ast\ast}$ \\ 
		~ & SCAMPI L1W$^\text{a}$	 			 & 0.2646 ± 0.0079 $^{\ast\ast}$	& 30.41 ± 0.34 $^{\ast\ast}$ & 0.8371 ± 0.0076 $^{\ast\ast}$ \\ 
		~ & SCAMPI TV$^\text{b}$		 & \textbf{0.2173 ± 0.0088} 	& \textbf{32.26 ± 0.41} & \textbf{0.8765 ± 0.0078} \\ 
		~ & SCAMPI L1W$^\text{b}$		 & 0.2301 ± 0.0095 $^{-}$	& 31.78 ± 0.41 $^{-}$ & 0.8647 ± 0.0084 $^{-}$ \\
		\midrule
		8 & PICS							 & 0.400 ± 0.023 $^{\ast\ast}$	& 27.00 ± 0.36 $^{\ast\ast}$ & 0.7620 ± 0.0099 $^{\ast\ast}$ \\ 
		~ & ENLIVE							 & 0.427 ± 0.011 $^{\ast\ast}$	& 26.22 ± 0.22 $^{\ast\ast}$ & 0.7800 ± 0.0065 $^{\ast\ast}$ \\ 
		~ & ConvDecoder 					 & \textbf{0.311 ± 0.010}	& \textbf{29.08 ± 0.37} & 0.7963 ± 0.0097 $^{\ast\ast}$ \\ 
		~ & U-Net						     & 0.410 ± 0.010 $^{\ast\ast}$   &   26.55 ± 0.26 $^{\ast\ast}$  &	0.7394 ± 0.0084 $^{\ast\ast}$ \\		
		~ & SCAMPI TV$^\text{a}$					 & 0.367 ± 0.010 $^{\ast\ast}$ & 27.57 ± 0.28 $^{\ast\ast}$ & 0.8033 ± 0.0076 $^{\ast\ast}$ \\ 
		~ & SCAMPI L1W$^\text{a}$	& 0.373 ± 0.010 $^{\ast\ast}$	& 27.41 ± 0.28 $^{\ast\ast}$ & 0.7907 ± 0.0081 $^{\ast\ast}$\\ 
		~ & SCAMPI TV$^\text{b}$				 & 0.315 ± 0.010 $^{-}$	& 28.94 ± 0.33 $^{-}$ & \textbf{0.8278 ± 0.0084} \\ 
		~ & SCAMPI L1W$^\text{b}$			 & 0.330 ± 0.010 $^{-}$	& 28.54 ± 0.32 $^{-}$ & 0.8144 ± 0.0088 $^{-}$ \\
		\bottomrule
	\end{tabular*}
\end{table*}

\subsection{Single coil imaging}\label{sec:results_1coil}
\begin{figure*}[tbh]
	\centering
	\includegraphics[width=0.95\textwidth]{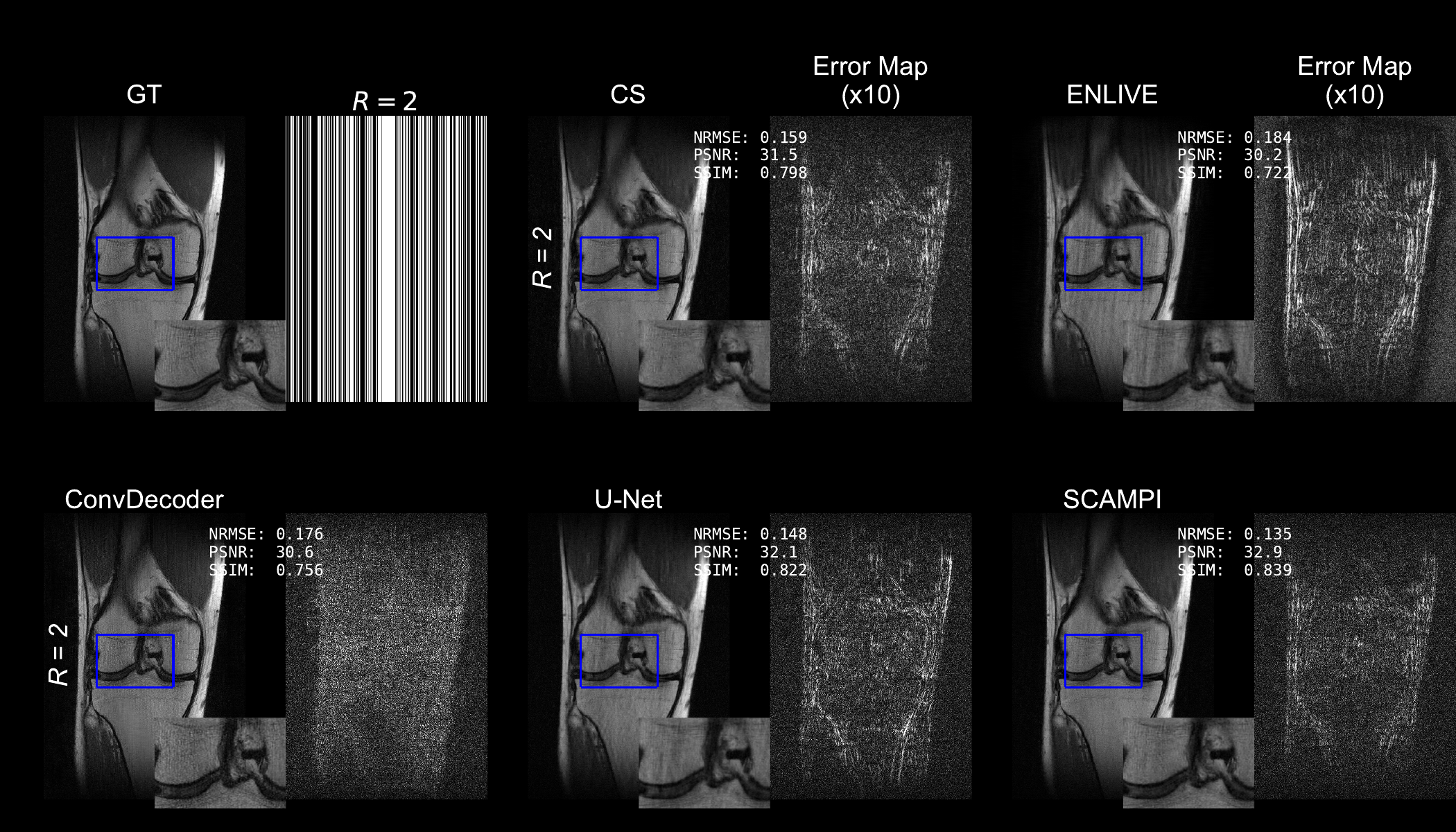}
	\caption{
		Undersampled single coil reconstruction ($R=2$) for SCAMPI (L1W regularization, $1000$ iterations) and reference methods ENLIVE, CS (L1W-regularization), \mbox{Conv}Decoder ($20000$ iterations) and U-Net with MSE penalty ($1000$ iterations).
		Quantitative image metrics NRMSE, PSNR and SSIM are reported next to the images. Relative difference (x10) to Ground Truth (GT) is shown right to each reconstruction.
		Inlay image shows zoom-in of the knee joint region (blue frame).
		\label{fig:results_1coil_knee}
	}
\end{figure*}

We studied the reconstruction of single coil MR data with SCAMPI and compared it with CS, ENLIVE and \mbox{Conv}Decoder. 
Hyperparameters for the reconstruction are reported in \secref{sec:methods}. 
The resulting reconstructions are illustrated in \figref{fig:results_1coil_knee} on an arbitrary example from the 200 samples subset described in \secref{sec:methods_single_coil_data}.
The reconstructions at ${R=3}$ are not presented, because none of the methods could produce high-quality reconstructions in the single coil case.

The results in \figref{fig:results_1coil_knee} demonstrate that even for ${R=2}$ ENLIVE is ineffective in reducing undersampling artifacts, when used with a single coil.
In the CS reconstruction, the undersampling artifacts are better suppressed, but the residual error indicates over-smoothing of the reconstruction. 
On the other hand, \mbox{Conv}Decoder yields visually appealing reconstructions that appear to remove undersampling artifacts. However, analysis of the error map reveals more random noise, than in the other reconstruction, which leads to inferior image metrics.
This matches the observations for the multicoil reconstructions, where the reconstruction noise maps for \mbox{Conv}Decoder also showed higher noise amplification than for the other Neural Network methods (cf.~\secref{sec:results_multicoil_noise}).
In contrast, SCAMPI exhibited the best reconstruction outcomes at ${R=2}$ (with similar results for TV and L1W regularization, which is why only the L1W results are illustrated here).

\paragraph{Quantitative analysis}
Quantitative analysis was performed on a subset of $200$ samples from the FastMRI singlecoil knee dataset, as described in \secref{sec:reconstruction-data}.
Reconstruction was performed with the parameter settings reported in \secref{sec:implementation-of-algorithms}.
Tab.~\ref{tab:stats_singlecoil} demonstrates, that the best reconstruction metrics are achieved by the SCAMPI algorithm.
\begin{table*}
	\centering
	\caption{	
		Mean image quality metrics for CS (L1W regularization), ENLIVE, ConvDecoder and SCAMPI (L1W regularization) on a random subset of $200$ samples from the FastMRI singlecoil knee dataset.
		Bold values highlight the method with best metrics for each $R$. Asterisks indicate significance level of differences between the respective method and the best one, 
		determined by a Mann-Whitney U rank test:  $^{-}: p > 0.4$, $^{\ast}: p < 0.05$, $^{\ast\ast}: p < 0.001$.
		Error margins denote the $95\%$ confidence interval.
		\label{tab:stats_singlecoil}
	}
	\begin{tabular*}{.85\textwidth}{@{\extracolsep\fill}|l|l|ccc|@{\extracolsep\fill}}
		\toprule
		\toprule
		R & Method & NRMSE & PSNR & SSIM \\ 
		\toprule
		2 & CS          & 0.264 ± 0.013  $^{\ast\ast}$ & 30.34 ± 0.57 $^{\ast\ast} $ & 0.735 ± 0.017 $^{\ast\ast}$  \\ 
		~ & ENLIVE      & 0.280 ± 0.012  $^{\ast\ast}$ & 29.75 ± 0.55 $^{\ast\ast} $ & 0.66 ± 0.014 $^{\ast\ast}$  \\ 
		~ & ConvDecoder & 0.297 ± 0.017  $^{\ast\ast}$ & 29.49 ± 0.66 $^{\ast\ast} $ & 0.677 ± 0.020 $^{\ast\ast}$ \\ 
		~ & U-Net		& 0.354 ± 0.022	 $^{\ast\ast}$ & 28.04 ± 0.68 $^{\ast\ast} $ & 0.660 ± 0.023 $^{\ast\ast}$ \\
		~ & SCAMPI TV   & \textbf{0.240 ± 0.013} & \textbf{31.30 ± 0.64} & \textbf{0.774 ± 0.015} \\ 
		~ & SCAMPI L1W  & 0.242 ± 0.013  $^{-}$ & 31.22 ± 0.62 $^{-}$ & 0.771 ± 0.015 $^{-}$ \\ 
		\midrule
		3 & CS		    & 0.331 ± 0.012 $^{\ast\ast}$ & 28.22 ± 0.49 $^{\ast\ast}$  & 0.639 ± 0.020 $^{\ast\ast}$ \\ 
		~ & ENLIVE      & 0.337 ± 0.012 $^{\ast\ast}$ & 28.08 ± 0.50 $^{\ast\ast}$  & 0.575 ± 0.017 $^{\ast\ast}$ \\ 
		~ & ConvDecoder & 0.361 ± 0.017 $^{\ast\ast}$ & 27.65 ± 0.60 $^{\ast\ast}$  & 0.579 ± 0.023 $^{\ast\ast}$ \\ 
		~ & U-Net		& 0.450 ± 0.022	$^{\ast\ast}$ & 25.75 ± 0.56 $^{\ast\ast}$  & 0.543 ± 0.025 $^{\ast\ast}$ \\		
		~ & SCAMPI TV   & 0.293 ± 0.014 $^{-}$ & \textbf{29.42 ± 0.57} & \textbf{0.685 ± 0.020} \\ 
		~ & SCAMPI L1W  & \textbf{0.302 ± 0.012} & 29.09 ± 0.52 $^{-}$ & 0.680 ± 0.019 $^{-}$ \\ 
		\bottomrule
	\end{tabular*}
\end{table*}

\section{Discussion and Conclusion}\label{sec:conclusions}
The application of SCAMPI for reconstruction of undersampled k-space data was demonstrated for 2D images from the FastMRI dataset.
Our approach requires no training database. 
It exploits a CNN's inherent capabilities for image reconstruction and generation without introducing the downsides of database-driven machine learning applications, such as limitations in generalizability and transformability~\cite{Darestani.2021b}
More importantly, it mitigates the risk of introducing biases from the training data into the reconstruction, which can cause hallucinating or leaving out image features.
Scan-specific training of Neural Networks has been investigated by other groups for mapping of $\mathbf{k}_0$ to $\mathbf{k}$
\cite{PourYazdanpanah.2019, Akcakaya.2019, Wang.2020}.
In contrast, SCAMPI uses the capabilities of a DIP for image generation, as illustrated by Ulyanov et al., and transfers it to a k-space reconstruction setting. Here, the reconstruction is searched in the parameter space of the Neural Network, which serves as a prior for preferable solutions. 
A DIP-based approach is also pursued by other groups \cite{DiZhao.2021, Shen.2022}.
However, they use previously measured reference data to pretrain the DIP model's network parameters before beginning the actual reconstruction task. 
This implies, that the actual reconstruction task cannot be considered as a \textit{de novo} image synthesis from the network parameters and only subject to matching $\mathbf{k}_0$.
Instead, it also depends on the choice of reference image.
Hence, this approach is not strictly free -- rather reduced -- of training-data other than the scan-data. It does not eliminate the above-mentioned risks of Neural Networks for bias towards training data.

A reference-free DIP approach was investigated in time-dependent radial MR-data by Yoo et al.~\cite{Yoo.2021}. By using latent mapping combined with generative CNNs, they achieved promising results with time-dependent MR-datasets.
Time series of MR-data, in general, show much higher informational redundancy as single images and thus are very suitable
for high acceleration rates \cite{feng2013highly,jung2009k,tsao2012mri}.
2D data, like those studied here, lack redundancies that are present in volumetric or time series data, which can impede reconstruction of highly undersampled measurements.
Therefore, other strategies for successful reconstruction are particularly interesting and could potentially be transferred to those scenarios as well.
We examined the reconstruction of 2D MRI with Cartesian undersampling with a U-Net as statistical prior and an enhanced loss function that includes sparsity regularization terms. 
Judged by different quantitative image metrics and visual artifact reduction, our method outperforms previously published approaches CS, ENLIVE~\cite{Holme.2019} and ConvDecoder~\cite{Darestani.2021} as well as a U-Net without our SCAMPI-loss term.
While all investigated Neural Network methods deliver better artifact reduction than the classical methods, SCAMPI brings an additional benefit by producing less noisier reconstructions, which we quantified with a Monte Carlo-based estimate of reconstitution noise.
A quantitative analysis of $200$ images from the FastMRI dataset showed the consistency of better SCAMPI reconstruction results in almost all investigated modalities.
In general, larger image errors were observed on this subset. We used reconstruction hyperparameters that were heuristically found during testing on a single image.
Although SCAMPI already outperformed reference methods for this direct approach, 
this could hint to an additional benefit that could come with optimizing hyperparameters to specific scan properties or with
further exploration of more general parameter settings on larger samples with variable imaging parameters.

Additionally, we assessed the reconstruction of undersampled single coil measurements.
In this case, the matrix inversion in the data consistency term of the CS optimization problem is underdetermined.
This leads to strong residual artifacts in the CS and ENLIVE reconstructions. 
Suppressing them with a heavily weighted sparsity-regularization term in the CS reconstruction leads to over-smoothed reconstructions that lack detailed structures.
In contrast, the untrained neural networks allow reconstructions with better preservation of fine structures. 
We attribute this to the additional benefit of the image prior, that is utilized.
Therefore, this approach is particularly interesting for accelerating single coil imaging where multichannel sampling is not feasible, for example in preclinical laboratories.
However, in the ConvDecoder reconstructions, there is more noise present, than in the other reconstructions.
It is described that early stopping of DIP can be used for denoising, while longer training leads to overfitting to noise from the corrupted training target~\cite{Ulyanov.2018, Wang.2021}.
Therefore, we assume that when interpolating k-space data at high undersampling rate with many iterations, ConvDecoder may hallucinate high-frequency noise.
Our method approaches denoising through early stopping, as we require fewer iterations.
This is supported by our analysis in \secref{sec:results_1coil}, which shows that SCAMPI delivers reconstructions with higher SNR than ConvDecoder.

In their work on DIP, Ulyanov et al. illustrated that CNNs are appropriate priors for natural images \cite{Ulyanov.2018}.
They allow the recovery of non-sampled image information due to the statistical redundancy that lies in spatial structure and correlations of images. 
Our approach utilizes these advantages by reconstructing $\mathbf{k}_0$ in the image domain while imposing both data fidelity in k-space and domain-transform sparsity, as known from CS.
Prevailing acceleration strategies use information redundancy in 3D, temporal or multicoil data. Usage of DIP extends these strategies with convolutional networks as priors for natural images.
Our approach bases on DIP and combines it with an extended loss function, including  sparsity-regularization to achieve higher reconstruction quality.
Since the Neural Network architecture is crucial for this task, future research on new approaches are of considerable interest.

As for now, a limitation of our work is the increase in reconstruction time compared to PICS. On our system, $12$-coil SCAMPI~(TV) takes approx. $98\,\mathrm{s}$ per reconstruction vs. $2\,\mathrm{s}$ for PICS~(TV).
However, due to the reduction of required iterations, it is considerably faster than ConvDecoder, which takes approximately $21\,\mathrm{min}$.
Moreover, the PICS algorithm is well-established and optimized, whereas
our DIP implementation is still at an experimental state with room for development. For single coil reconstruction, we observed a computation time of approx. $90\,\mathrm{s}$. The reduction from $98\,\mathrm{s}$ does not scale with the reduction in computational effort from ${n_c=12}$ to ${n_c=1}$.
This hints at potential for code-optimization, which is beyond the scope of this work.
In addition, our selection of $1000$ iterations was merely chosen as a pragmatic trade-off between reconstruction quality and time efficiency. The non-linear loss-curves during iteration (cf. Suppl. materials) indicate that acceptable results could as well be achieved with, e.g., half the number of iterations, which might deserve future exploration.
Also, future development of advanced hardware is to be expected and will allow faster and more extensive computations in image reconstruction.
Moreover, the anatomical variance of our samples is limited on basal slices of axial brain images and central slices of knee scans. Although \figref{fig:contrasts} shows that there is still ad-hoc variance in the selected samples, an extension to further anatomical regions could be of interest to validate whether our method does generalize well.
Additionally, further research is necessary, to estimate robustness and uncertainty of reconstruction results with untrained neural networks \cite{Antoran.2022}. 

SCAMPI uses contemporary a tensor network and stands to gain from future discoveries on refined network architectures, training, etc..
For example, the random input variables $\mathbf{z}$ could also be trained scan-specifically to better fit the desired output $\mathbf{x}$. However, preliminary experiments on trainable inputs showed only minor difference. This could be attributed to the fact that the parameter space is only marginally increased by trainable inputs: the calibrated $n_c=12$ SCAMPI network for a complex target image of $396^2$ pixels and a trainable input layer has about $17.5\times10^6$ parameters as opposed to $17.3\times10^6$ without trainable inputs. Since the over-parameterization of $\mathbf{x}$ by the network plays a key role in deep image prior
\cite{Ulyanov.2018,Hand.2019}.

Finally, it should be pointed out that artifact correction by sparsity regularization requires incoherent `noise-like' artifacts \cite{lustig2007sparse, Candes.2006}. 
Therefore, as for CS,
better performance can be expected for less coherent k-space trajectories, for example with Poisson, radial or spiral sampling.
In addition, 3D or spatio-temporal data, that bring further redundancy in the acquired data should allow for higher acceleration rates.
Applying our approach to these modalities can be a promising field of future research.

Compared to reference methods with untrained neural networks, SCAMPI offers a fast and robust way of k-space interpolation.
It opens the field to various new possibilities for further improvements of network architecture and hyperparameters without introducing the downsides of learning from databases.

\section*{Acknowledgments}
The authors want to thank Peter Dawood for many discussions on related topics.
This work was partially supported by Deutsche Forschungsgemeinschaft (DFG) grant 396923792. 

\subsection*{Financial disclosure}

None reported.

\subsection*{Conflict of interest}

The authors declare no potential conflict of interests.

\subsection*{Code and data availability}
Python code for SCAMPI and example data are available on
\href{https://github.com/vrherold/Scampi}{github.com/vrherold/Scampi}.
The datasets used and/or analyzed during the current study are available from the corresponding author on reasonable request.

\subsection*{Supplementary Information}
Supplementary information accompanies this paper. 


\vfill\pagebreak

\end{document}